# Electrical conductivity in extremely disordered molybdenum oxynitrides thin films

J. A. Hofer,[1*] S. Bengio,[2] G. Rozas,[1,2] P. D. Pérez,[2] M. Sirena,[1,2] S. Suárez,[1,2] N. Haberkorn.[1,2]

[1] *Instituto Balseiro, Universidad Nacional de Cuyo and Comisión Nacional de Energía Atómica, Av. Bustillo 9500, 8400 San Carlos de Bariloche, Argentina.*

[2] *Comisión Nacional de Energía Atómica and Consejo Nacional de Investigaciones Científicas y Técnicas, Centro Atómico Bariloche, Av. Bustillo 9500, 8400 San Carlos de Bariloche, Argentina.*

We report on the influence of the chemical composition on the electronic properties of molybdenum oxynitrides thin films grown by reactive sputtering on Si (100) substrates at room temperature. The partial pressure of Ar was fixed at 90 %, and the remaining 10 % was adjusted with mixtures $N_2:O_2$ (varying from pure $N_2$ to pure $O_2$). The crystalline and electronic structures and the electrical transport of the films depend on the chemical composition. Thin films grown using oxygen mixtures up 2 % have γ-$Mo_2N$ phase and display superconductivity. The superconducting critical temperature $T_c$ reduces from ~ 6.8 K to below 3.0 K as the oxygen increases. On the other hand, films grown using oxygen mixtures richer than 2 % are mostly amorphous. The electrical transport shows a semiconductor-like behavior with variable-range hopping conduction at low temperatures. The analysis of the optical properties reveals that the samples have not a defined semiconductor band gap, which can be related to the high structural disorder and the excitation of electrons in a wide range of energies.

## *1. Introduction*

The structural, mechanical and electronic properties of ceramics usually depend on the doping level and the type of anions. For oxynitrides, the oxidation state of the cations, the bond covalence and the band structure can be tuned by mixing oxygen and nitrogen. The wide range of electronic and optical properties make them promising candidates for photocatalysis [1,2], electronic [3,4,5] and solar energy applications [6]. Depending on the potential application, the study of the physical properties of metallic oxynitrides usually is performed in nanoparticles [7] and in thin films [8]. In addition to the doping, the properties of the materials are also affected by the microstructure. Among the chemical and physical methods for the fabrication of thin films, sputtering has the advantage that it is possible to obtain from amorphous to epitaxial structures by modifying the substrate temperature.

From an electronic point of view, it is interesting to study the role of the oxygen doping on the properties of superconducting nitrides. For materials such as TiN [9] and NbN [10], the structural and electrical properties can be tuned adding oxygen and other dopant elements [11,12]. For example, adding oxygen a crossover from a dirty superconducting nitride (due to paramagnetic oxygen impurities) to a semiconductor oxynitride can be expected [13]. Among the candidates to analyze the effect of stoichiometry on the electrical and structural properties are molybdenum oxynitrides, The most common molybdenum oxides are the monoclinic dioxide ($MoO_2$) and the trioxide ($MoO_3$). $MoO_2$ exhibits unusual properties among oxides, shifting from one with metallic properties into a semiconductor when the structural disorder increases. [14,15,16,17]. $MoO_3$ is *n*-type semiconductor [18]. On the other hand, molybdenum nitrides present several superconducting crystalline phases: $\gamma$-$Mo_2N$ (cubic) with a superconducting critical temperature $T_c$~ 5 K [19], $\beta$-$Mo_2N$ (tetragonal) with $T_c$ ~ 5 K [20] and $\delta$-MoN (hexagonal) with $T_c$ ~ 12 K [21,22]. The chemical composition of molybdenum nitride thin films grown by reactive sputtering can be adjusted by the gas mixture [23]. A distinctive property of $\gamma$-$Mo_2N$ thin films is that its $T_c$ rises from ≈ 5 K to 8 K as the disorder at the nanoscale increases [24,25]. Notwithstanding the rich properties displayed by the nitrides and oxides, there are only a few studies on the electrical properties of molybdenum oxynitrides [26,27].

In this work, we analyze the influence of the reactive gas mixture on the structural, chemical and electronic properties of molybdenum oxynitrides thin films grown at room temperature by reactive sputtering on Si (100) substrates. The reactive atmosphere is an Ar:$N_2$:$O_2$ mixture. The Ar: ($N_2$+$O_2$) ratio was fixed in 90:10 and the $N_2$:$O_2$ ratio varied from 0 to 10. The results show that the microstructure and electrical properties depend on the gas mixture. Nitrogen-rich films are superconducting and the $T_c$ is gradually reduced by increasing the oxygen doping. Oxygen-rich films display a semiconductor-like behavior. The physical properties of the samples are analyzed by considering the influence of the chemical doping and of the structural disorder.

## 2. *Material and methods*

Molybdenum oxynitride films were deposited by DC reactive magnetron sputtering on Si (100) (typical size 1 cm$^2$). No intentional heating of the substrate was used. The base pressure in the chamber was 1.3x10$^{-4}$ Pa. Films were grown from a pure Mo target (diameter 3.3 cm) in a reactive Ar:$N_2$:$O_2$ mixture. The target power was fixed at 50 W and the total pressure in 0.67 Pa. The substrate was positioned directly over the target at ≈ 5.5 cm. Reactive sputtering was performed with 90% Ar and 10 % of a $N_2$:$O_2$ mixture (with $O_2$ total fraction 0, 1%, 2%, 3.3%, 5%, 6.6% and 10%). The notation [MoN$_x$O$_y$] indicates thin films growth in a $N_2$:$O_2$ gas mixture where *x* and *y* are the respective percentages.

X-ray (XRD) diffraction data was obtained using a Panalytical Empyrean equipment operated at 40 kV and 30 mA with the Cu$_{K\alpha}$ radiation. The structural analysis was performed based on Θ-2Θ scans with an angular resolution of 0.02°. The film thickness was measured by low-angle X-ray reflectivity (XRR). AFM measurements were performed in a Dimension 3100 ©Brucker microscope. The AFM images presented in this work were performed in tapping mode. The chemical stoichiometry of the films was analyzed by Rutherford Backscattering Spectroscopy (RBS) with a TANDEM (NEC, 1.7 MV) accelerator using a 2 MeV $^4$He$^{2+}$ ion beam. Surface composition analysis was performed by X-ray photoelectron spectroscopy (XPS) using a standard Al/Mg twin-anode X-ray gun and a hemispherical electrostatic electron energy analyzer (high vacuum conditions with a base pressure of 10$^{-9}$Torr).

The optical parameters of for [MoN$_{6.6}$O$_{3.3}$], [MoN$_{3.3}$O$_{6.6}$], and [MoN$_0$O$_{10}$] were deduced from spectroscopic ellipsometry measurements (Ψ and Δ) carried out in the spectral region of 200–1200 nm using a Woollam NIR-Vis-UV ellipsometer in three different angles of incidence: 61°, 66° and 71°. The data was analyzed using the manufacturer's code. The electrical transport measurements were performed using the standard four-point configuration.

## 3. Results and discussion

The thicknesses of the films were determined from XRR. The modulation in the Θ-2Θ scans is related to the thickness *d* of the film as:

$$sin^2\theta = [\frac{(n+K)\lambda}{2d}]^2 + 2\delta, \quad [1]$$

where 1-δ is the real part of the index of refraction of the film, and *k* = 0 (intensity minimum) and *k* = ½ (intensity maximum) [28]. Figure 1*a* shows $sin^2\theta - vs - n^2$ for the low angle minima for films grown for 5 minutes using different gas mixtures. Inset Fig. 1*a* shows the XRR data for [MoN$_{10}$O$_0$]. Figure 1*b* shows a summary of the total thickness. The results indicate that the growth rate using Ar: N$_2$ is 16 nm/ min (see Inset Fig.1*b*). The value rises with a clear crossover at ≈ 2% O$_2$ as the oxygen in the mixture increases. Finally, the growth rate for ≈ 10% O$_2$ (no N$_2$) is 28 nm/min. The analysis of the surface topology indicates that is weakly affected by the gas mixture. Figure 2 shows AFM images for the extremes [MoN$_{10}$O$_0$] and [MoN$_0$O$_{10}$]. The films display very smooth surfaces with Root Mean Square (RMS) roughness smaller than 0.5 nm.

Figure 3 shows the XRD patterns for molybdenum oxynitride films. Nitrogen-rich films display reflections corresponding to the γ-Mo$_2$N phase [23]. For pure N$_2$ and low O$_2$ concentrations, the films are textured along the (200). The reflections (111) and (200) are observed in [MoN$_8$O$_2$]. Gas mixtures richer than 2 % in O$_2$ produce mainly amorphous films, which is evident from the absence of peaks in the XRD patterns. The change in the microstructure from nanocrystalline Mo$_2$N phase to an amorphous phase is coherent with the increment in the growth rate described earlier. To understand about the influence of the stoichiometry on the structural changes, we study the chemical composition by RBS. Table 1 shows a summary of the results. The error bar for the data is 5 %. The extremes [MoN$_{10}$O$_0$] and [MoN$_0$O$_{10}$] correspond to Mo$_2$N$_{1.1}$ and MoO$_2$, respectively. The data show that for gas mixtures with oxygen above 2% the stoichiometry of nitrogen drops and the samples are mainly oxides. This crossover in the chemical composition is in agreement with the presence of amorphous structures in the XRD data.

XPS analysis was employed to investigate the electronic structure of the pristine and cleaned films. The surface cleaning was performed with Ar$^+$ sputtering (2 kV). The pristine films display a component at binding energy (BE) ≈ 232.7 eV related to superficial $MoO_3$ that is removed during the sputtering process [29]. Figure 4 shows the XPS spectra in the Mo$3d$ region for cleaned [MoN$_{10}$O$_0$], [MoN$_8$O$_2$], [MoN$_5$O$_5$] and [MoN$_0$O$_{10}$]. The results can be divided into nitrides and oxides. The spectra for cleaned [MoN$_{10}$O$_0$], and [MoN$_8$O$_2$], display the Mo$_2$N component (binding energy BE ≈ 228.5 eV) shifted to smaller BE by 0.2 eV (see Fig. 4a). The electronic structure for oxygen mixtures richer than 2 % displays the $MoO_2$ with BE ≈229.3 eV (not shown). After clean the surface, [MoN$_5$O$_5$] and [MoN$_0$O$_{10}$] suffer a drastic reduction showing MoO and metallic Mo (see Fig.4$b$) [30,31,32].

To understand in more detail the electronic structure of the films, we analyzed its valence band spectra. Figure 5 shows a summary of the results. The comparison between [MoN$_{10}$O$_0$] and [MoN$_8$O$_2$] indicates that for the same contribution of N2s, the first displays a higher O2s intensity (see Fig. 5$a$). [MoN$_8$O$_2$] displays a larger contribution of the metallic Mo$4d$. The latter is also observed at [MoN$_5$O$_5$] and [MoN$_0$O$_{10}$] (see Fig. 5$b$). The contribution of N drops as the oxygen increases. It is important to note that nitrides are more stables than oxides to the sputtering process [33]. No nitrogen vacancies are formed during the sputtering process used to clean the surface in [MoN$_{10}$O$_0$]. While oxygen vacancies are formed in [MoN$_5$O$_5$] and [MoN$_0$O$_{10}$]. The spectrum for [MoN$_8$O$_2$] suggests that the sample is affected by the creation of oxygen and presumably nitrogen vacancies. This supposition explains the shift in the shift in the Mo3d spectra and the rise in the Mo4d intensity relative to the O1s and N1s. Furthermore, the width of the Mo4d peak and its proximity to the N2p orbital reveals a high hybridization (see top panel Fig. 5), while the narrow Mo4d peak and its distancing with the O2p orbital indicate lower hybridization (see bottom panel Fig. 5).

To correlate the microstructure and the electronic properties, we measured the electrical resistivity versus temperature for the different films (see Fig. 6). The results show that films with reactive mixtures with O$_2$ up 2% display superconductivity. The $T_c$ decreases from ≈ 6.8 K for [MoN$_{10}$O$_0$] to ≈ 3 K for [MoN$_8$O$_2$]. Thin films grown using oxygen mixtures above 3.3 % display a semiconductor-like behavior. The measured resistivity at 273 K was found to $\rho^{273 K}$ = 100 (20) μΩ.cm for [MoN$_{10}$O$_0$], 140 (20) μΩ.cm for [MoN$_9$O$_1$], 180 (20) μΩ.cm for [MoN$_8$O$_2$], 200 (20) μΩ.cm for [MoN$_{6.6}$O$_{3.3}$], 250 (20)

μΩ.cm for [MoN$_5$O$_5$], 700 (20) μΩ.cm for [MoN$_{3.3}$O$_{6.6}$], and 1500 (50) μΩ.cm for [MoN$_0$O$_{10}$],

Following, the influence of the oxygen impurities in the superconducting properties is analyzed by measuring the upper critical field ($H_{c2}$). The temperature dependence of $H_{c2}$ for dirty superconductors is described by the Werthamer-Helfand-Hohenberg (WHH) formula [34]:

$$ln\frac{1}{t} = \sum_{v=-\infty}^{\infty}\left(\frac{1}{|2v+1|} - \left[|2v+1| + \frac{\hbar}{t} + \frac{(\alpha\hbar/t)^2}{|2v+1|+(\hbar+\lambda_{so})/t}\right]^{-1}\right), \quad [2]$$

where $t = T / T_c$, $\hbar = (4/\pi^2)(H_{c2}(T)/|dH_{c2}/dT|_{T_c})$, α is the Maki parameter, and $\lambda_{so}$ is the spin-orbit scattering constant. When $\lambda_{so} = 0$, $H_{c2}(0)$ obtained from the WHH formula satisfies the relation $H_{c2}(0) = \frac{H_{c2}^{orb}(0)}{\sqrt{1+\alpha^2}}$ [35]. Figure 7 shows the summary of the results. The inset in Fig. 7 shows typical curves of normalized resistance versus temperature for different magnetic fields in [MoN$_9$O$_1$]. The experimental data is well described by the WHH model using α = 0 and $\lambda_{so}$ = 0 (see dashed lines in Fig. 7). The obtained $H_{c2}$ (0) values are 12 T for [MoN$_{10}$O$_0$], 9 T for [MoN$_9$O$_1$] and 6 T for [MoN$_8$O$_2$]. The coherence length ξ (0) values can be estimated using $\xi(0) = \sqrt{\Phi_0/(2\pi H_{c2}^{\|}(0))}$ (with $\Phi_0$ = 2.07 x 10$^{-7}$ G cm$^2$ is the flux quantum). The obtained ξ (0) values are 5.2 nm for [MoN$_{10}$O$_0$], 6 nm for [MoN$_9$O$_1$] and 7.4 for [MoN$_8$O$_2$]. It is important to note that for a weakly coupled BCS superconductor with similar band structure $\xi_0 = 0.18\frac{\hbar v_F}{k_B T_c}$ ($v_F$: Fermi velocity and $K_B$ the Boltzmann constant) [36]. The $\xi_0 * T_c$ value for the different films decreases as the oxygen increases, which suggests modifications in the band structure.

Now we will analyze the semiconductor-like behavior for [MoN$_5$O$_5$], [MoN$_{3.3}$O$_{6.6}$], and [MoN$_0$O$_{10}$] in more detail. In general, the temperature dependence of the resistivity in disordered systems and amorphous semiconductors takes the following form:

$$\rho \approx \rho_0 \exp\left[-\left(\frac{T_0}{T}\right)^p\right], \quad [3]$$

where $\rho_0$ is a prefactor, $T_0$ is a characteristic temperature and the exponent $p$ depends on the shape of the density of states at the Fermi level (FL) [37]. For Mott variable range hopping (VHR) $p$ can be ¼ (3D systems) or 1/3 (2D systems). Moreover, $p$ = ½ is expected for 2D systems in which the Coulomb interaction is important. Figure 8 shows $ln(\rho)$ vs. $T^{-1/4}$ for the analyzed samples. Straight lines are observed at low temperatures. To verify 3D VRH at low temperatures with $p$= 1/4, the equation [3] can be rewritten as $W = -\delta(ln\sigma(T))/\delta(lnT) = p(T/T_0)^p$ [38]. Inset Fig. 8 shows

$lnW$ versus $lnT$ for [MoN$_{3.3}$O$_{6.6}$], the slope $p$ = 0.24 (0.01) confirms the mechanism. The $T_0$ values obtained from the slopes are 1.4 K, 4.5 K and 25 K, for [MoN$_5$O$_5$], [MoN$_{3.3}$O$_{6.6}$], and [MoN$_0$O$_{10}$], respectively. From the values of $T_0$ it is possible to estimate the hopping energy $E_h(T)$ for a given temperature $T$ [39]:

$$E_h(T) = \frac{1}{4}k_B T^{3/4} T_0^{1/4}. \quad [4]$$

The $E_h$ (5 K) goes from 10 µeV to 160 µeV when the oxygen is increased. These values are much smaller than those usually observed for more insulating samples such as manganites and ZnO [39,40].

The semiconductor-like behavior in the samples was analyzed by ellipsometry measurements. Refractive index ($n$) and extinction coefficient (k) were calculated from the modeling of the elipsometric variables ($\Psi$ and $\Delta$). Figure 9a shows the results for wave lengths $\lambda$ between 200 nm and 1200 nm for [MoN$_5$O$_5$], [MoN$_{3.3}$O$_{6.6}$] and [MoN$_0$O$_{10}$]. The refraction index increases monotonically as the energy decreases. Moreover, K ($\lambda$) is different from the dependence expected for semiconductors with a defined band gap. There is not a crossover to lower absorption when the energy decreases. For amorphous semiconductors, there is not long range atomic order. However, the short-range order remains to some extent, giving rise thereby to a band-like structure of electron energy states similar to that of crystalline semiconductors. Nevertheless, the absorption edge becomes indistinguishable due to the high disorder. Electrons in the called diffuse band can contribute to the conduction even for low energies. To verify the presence of a gap when the disorder is reduced, [MoN$_0$O$_{10}$] was annealed at 600 °C for 1 hour using a vacuum of 1.3x10$^{-4}$ Pa (sample 1) and 101325 Pa O$_2$ (sample 2). Figure 9b shows the $n$ ($\lambda$) and the $K$ ($\lambda$) dependences after annealing. The latter displays a strong decrease in the visible and near-infrared (NIR) range in agreement with the expectations for semiconductors. In the case of the oxygen-annealed sample, we cannot estimate reliable optical properties because the small absorption and thickness of the annealed material difficult the fitting of the elipsometric optical model. In fact, the obtained $n$ and k values are greatly influenced by the optical properties of the Si substrate. For the vacuum-annealed sample, however, we can estimate a band gap using an $(\alpha h\nu)^2$ vs. ($h\nu$) plot (with $\alpha = 4\pi k/\lambda$) [16,41]. Inset Fig. 9b shows the obtained results for annealed [MoN$_0$O$_{10}$]. Two different band gaps can be identified. The presence of defined gaps indicates that the thermal annealing increases the order and reduces the contribution of the called diffuse band. The film annealed in vacuum display energy gap values of 2.15 eV and 2.74 eV. The values are in the range of those previously reported for $MoO_2$ and $MoO_3$ [16][41]. The presence of two gaps in the samples may be related to disorder and changes in oxygen stoichiometry. Similar features have been previously observed in thermal annealed Mo oxide thin films obtained by electrodeposition [16].

## 4. *Conclusions*

In summary, we analyzed the influence of the chemical composition on the electronic properties of molybdenum oxynitrides thin films grown by reactive sputtering on Si (100) substrates at room temperature. The electronic properties of the films are affected by the composition of the reactive gas mixture. For rich $N_2$, the films are superconducting. The $T_c$ is systematically reduced from $\approx$ 6.8 K for Mo-64 *at.*% N-36 *at.*% to $\approx$ 3 K for Mo-53 at.%N-30 *at.*% O-17 *at.*%. The oxygen is an interstitial impurity in the superconducting γ-Mo2N phase and is also segregated as amorphous $MoO_2$. For rich $O_2$ mixtures, the films are mainly amorphous oxide and display a semiconductor-like behavior. The electrical resistivity depends on the oxygen content. The electrical transport shows a semiconductor-like behavior with a VRH conduction at low temperatures. The analysis of the optical properties reveals that the samples have not a defined semiconductor band gap, which can be related to the high disorder and the excitation of electrons in a wide range of energies. The presence of a semiconductor gap is evidenced for annealed samples. Further investigations on the influence of the thermal annealing on the electronic should contribute to understand the role of the disorder on the resulting electronic properties.

**Acknowledgments**

This work was partially supported by the ANPCYT (PICT 2015-2171), U. N. de Cuyo 06/C505 and CONICET PIP 2015-0100575CO. JAH, SBM, GR, MS and NH are members of the of the Instituto de Nanociencia y Nanotecnología, CNEA-CONICET.

Declarations of interest: none.

Table 1. Chemical composition (atomic (%)) obtained from RBS measurements. The error bars for the data are estimated in approximately 5 %.

Figure 1. a) $sin^2\theta - vs - n^2$ for the minima in the XRR data. Solid line is the least-squares linear fit to the data. Inset: typical XRR data for [MoN$_{10}$O$_0$]. *b)* Thickness for films grown during 5 minutes using different gas mixtures. The data is calculated from the slope of *a)*. Inset: Growth rate for the different gas mixtures.

Figure 2. AFM topographical images (10×10 µm$^2$) of [MoN$_{10}$O$_0$] (left) and [MoN$_0$O$_{10}$] (right).

Figure 3. XRD for cleaned molybdenum oxynitride films.

Figure 4. XPS Mo3d spectra for [MoN$_{10}$O$_0$], [MoN$_8$O$_2$], [MoN$_5$O$_5$] and [MoN$_0$O$_{10}$].

Figure 5. XPS valence band spectra for surface cleaned [MoN$_{10}$O$_0$], [MoN$_8$O$_2$], [MoN$_5$O$_5$] and [MoN$_0$O$_{10}$]. Inset shows the spectra for pristine [MoN$_5$O$_5$] and [MoN$_0$O$_{10}$].

Figure 6. *a)* Electrical resistivity versus temperature for molybdenum oxynitride films grown using different gas mixtures. *b)* Normalized resistance vs. temperature at *T* < 10 K for superconducting samples. *c)* Summary of $T_c$ versus oxygen concentration in the reactive mixture.

Figure 7. Temperature dependence of the upper critical field (H$_{c2}$) for [MoN$_{10}$O$_0$], [MoN$_9$O$_1$] and [MoN$_8$O$_2$]. Inset shows typical curves of the resistance for different applied magnetic fields for [MoN$_9$O$_1$].

Figure 8. ln ρ vs $T^{1/4}$ for [MoN$_5$O$_5$], [MoN$_{3.3}$O$_{6.6}$] and [MoN$_0$O$_{10}$]. The inset shows ln *W* vs ln *T* for [MoN$_{3.3}$O$_{6.6}$] (with $W = -\delta(ln\sigma(T))/\delta(lnT)$).

Figure 9. Wavelength dependence of the refractive index (n) and extinction coefficient (k) of: *a)* as-deposited [MoN$_5$O$_5$], [MoN$_{3.3}$O$_{6.6}$] and [MoN$_0$O$_{10}$]; b) [MoN$_0$O$_{10}$] annealed at 600 °C in vacuum and under 101325 Pa O$_2$. Inset shows (αhν)$^2$ vs. hν curves for vacuum-annealed [MoN$_0$O$_{10}$].

Figure 1.

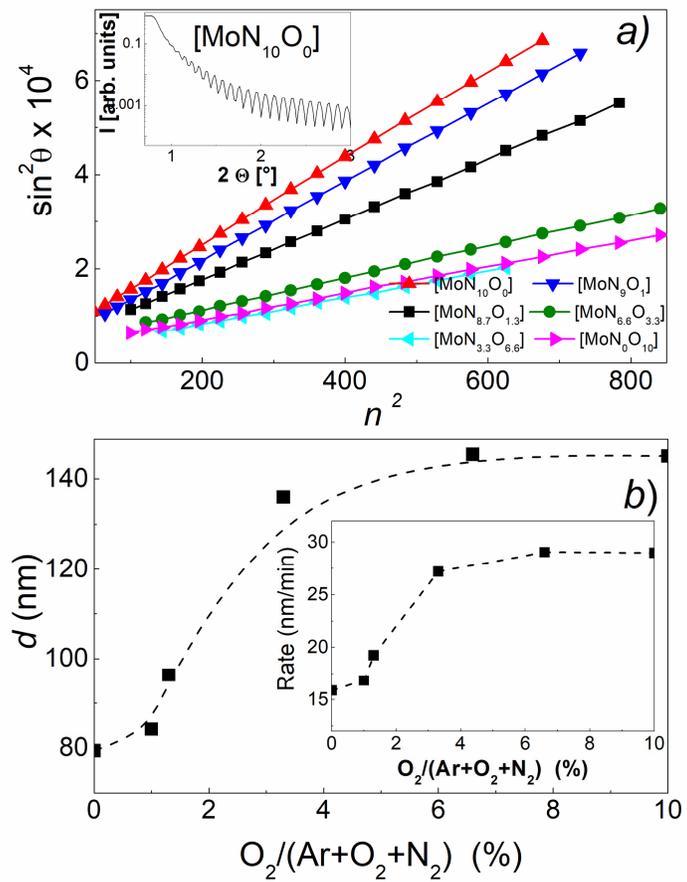

Figure 2.

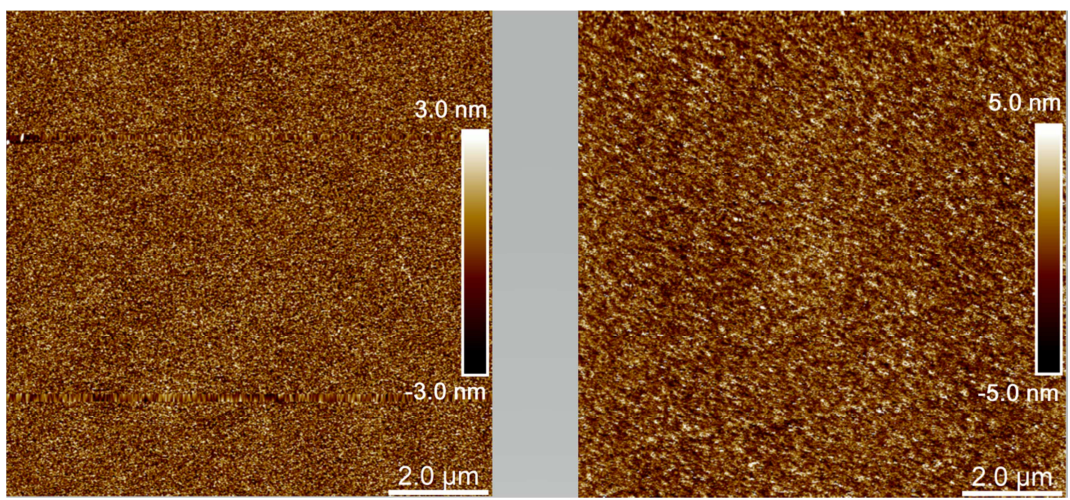

Figure 3.

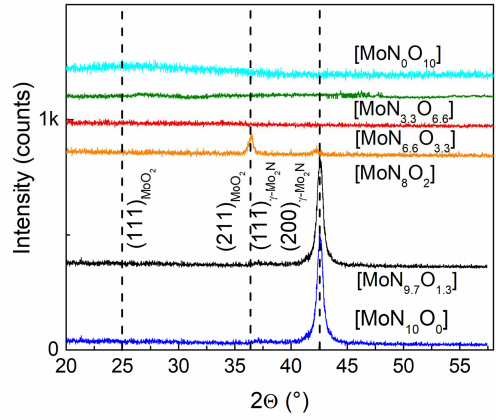

Figure 4.

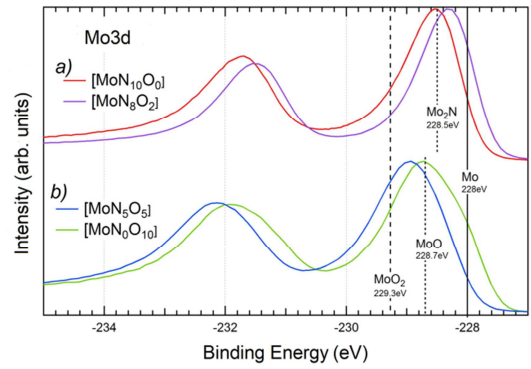

Figure 5.

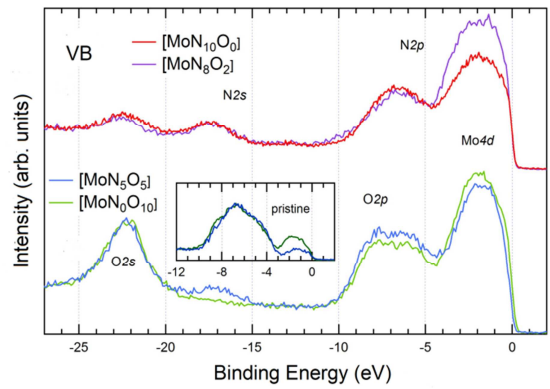

Figure 6.

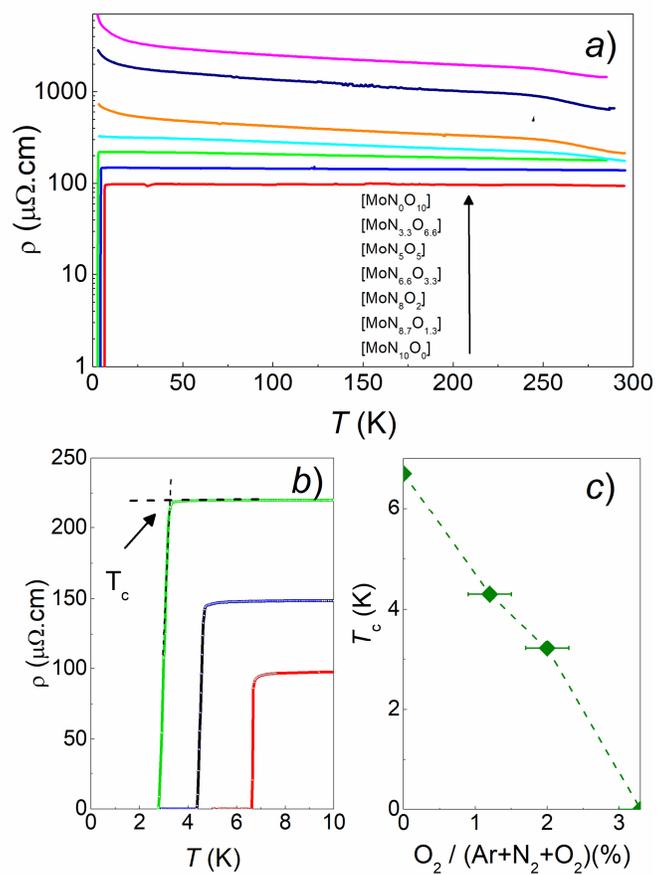

Figure 7.

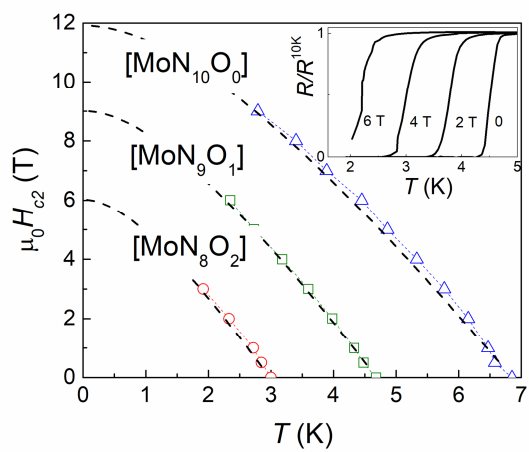

Figure 8.

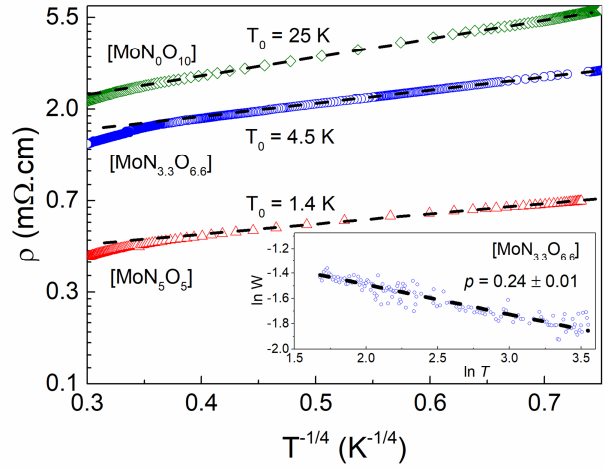

Figure 9.

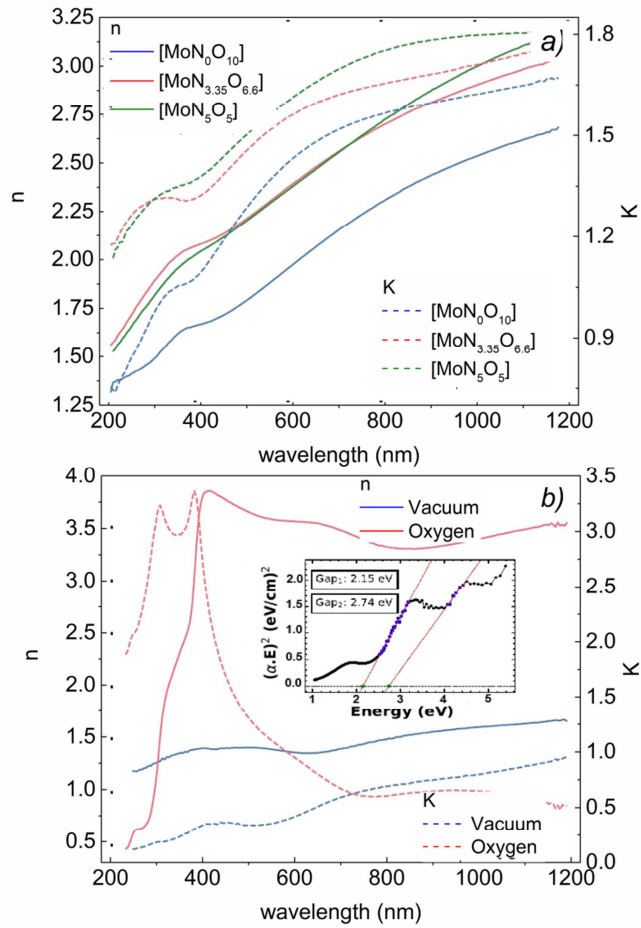

Table I.

| Sample | Mo | N | O |
|---|---|---|---|
| [MoN$_{10}$O$_0$] | 0.64 | 0.36 | -- |
| [MoN$_9$O$_1$] | 0.54 | 0.34 | 0.12 |
| [MoN$_8$O$_2$] | 0.53 | 0.3 | 0.17 |
| [MoN$_{6.6}$O$_{3.3}$] | 0.34 | 0.26 | 0.40 |
| [MoN$_5$O$_5$] | 0.40 | 0.24 | 0.36 |
| [MoN$_{3.3}$O$_{6.6}$] | 0.40 | 0.24 | 0.35 |
| [MoN$_0$O$_{10}$] | 0.33 | -- | 0.67 |